# Non-contact method for measurement of the microwave conductivity of graphene

L Hao[1,2], J Gallop[1], S Goniszewski[1,2], A Gregory[1], O Shaforost[2], N Klein[2] and R Yakimova[3]

[1]National Physical Laboratory, Hampton Road, Teddington, TW11 0LW, UK

[2] Imperial College London, South Kensington Campus, London, SW7 2AZ, UK

[3]Department of Physics, Chemistry and Biology (IFM), Linköping University, S-581 83 Linköping, Sweden

E-mail: ling.hao@npl.co.uk

**Abstract.** We report a non-contact method for conductivity and sheet resistance measurements of graphene samples using a high Q microwave dielectric resonator perturbation technique, with the aim of fast and accurate measurement of microwave conductivity and sheet resistance of monolayer and few layers graphene samples. The dynamic range of the microwave conductivity measurements makes this technique sensitive to a wide variety of imperfections and impurities and can provide a rapid non-contacting characterisation method. Typically the graphene samples are supported on a low-loss dielectric substrate, such as quartz, sapphire or SiC. This substrate is suspended in the near-field region of a small high Q sapphire puck microwave resonator. The presence of the graphene perturbs both centre frequency and Q value of the microwave resonator. The measured data may be interpreted in terms of the real and imaginary components of the permittivity, and by calculation, the conductivity and sheet resistance of the graphene. The method has great sensitivity and dynamic range. Results are reported for graphene samples grown by three different methods: reduced graphene oxide (GO), chemical vapour deposition (CVD) and graphene grown epitaxially on SiC. The latter method produces much higher conductivity values than the others.

## 1. Introduction

The remarkable properties of single- and few-layer graphene thin films has led to an explosion of activity [1-4]. A number of different methods for preparing graphene thin films have appeared and a wide range of experiments are being carried out on them [5-12]. There is a great deal of variability in the quality of films prepared, and even when identical methods are used the film properties between successive batches may be quite different. The accepted method for characterising the electrical properties of graphene films is to measure the mobility. However this generally requires patterning of the films and making electrical contact. The provision of a gate voltage to tune the carrier density is also often necessary. The additional processes mean that quality assessment of the films is time consuming and requires physical intervention on the graphene wafer. Here we report development of a quick, non-invasive and non-contacting method for measurement of the microwave surface impedance (and hence the conductivity) and sheet resistance of graphene thin films. A similar technique has been used previously to examine a variety of other materials [13, 14]. However the advantage of our method is that exact solution of the mode geometries is not necessary. Under the conditions which are specified below it is not necessary to carry out a detailed mode matching or finite element electromagnetic model to derive the electrical parameters of the TE film. The conductivity and sheet resistance are derived by reference to measurements on a similar bare substrate to that on which the graphene sample is deposited.

## 2. Outline of the method

Consider the arrangement shown in figure 1a. A single crystal sapphire puck acts as a microwave dielectric resonator, contained within a copper housing. It is spaced away from the copper housing by a short quartz spacer tube. A plain low loss dielectric substrate, of thickness $t_s$, can be brought to a fixed position in relation to this sapphire puck resonator (see figure. 1b) and the resulting shift in both the resonant frequency $\Delta f_s$ and the linewidth $\Delta w_s$ can be measured, provided the quality factor of the sapphire resonator is high enough compared with the losses contributed by the substrate. Now take another, nominally identical substrate, coated with a uniform layer of graphene of thickness $t_g$ (see figure 1c). Position it in the same position relative to the sapphire puck and make further measurements of the resonant frequency shift $\Delta f_g$ and linewidth shift $\Delta w_g$. Finally measure the unperturbed sapphire puck resonant frequency $f_0$ and linewidth $w_0$ with only puck and support quartz tubes in the copper housing (see figure 1d).

## 3. Quantitative Analysis

Since $t_g << t_s$ and also $t_g << \delta_g$, the electromagnetic skin depth of graphene at microwave frequencies we may assume, to a good approximation, that the field distributions in the bare substrate and graphene coated substrate situations are the same. Thus we may apply perturbation theory to evaluate the surface impedance of the graphene, provided that the complex permittivity and thickness of the bare substrate are known.

$$\Delta f_s = f_0 \left( \frac{(\varepsilon_s' - 1) \int E^2 dV}{W} \right) \quad (1)$$

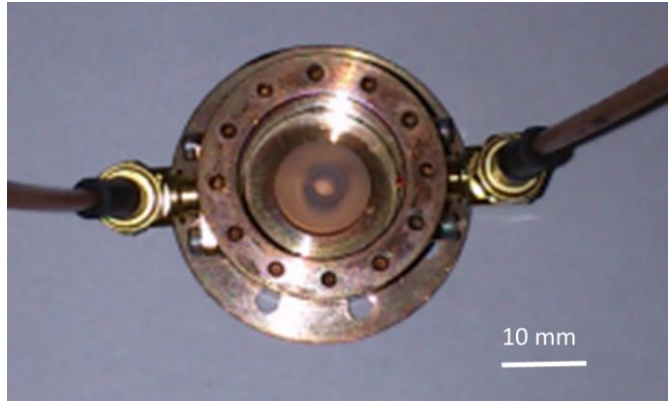

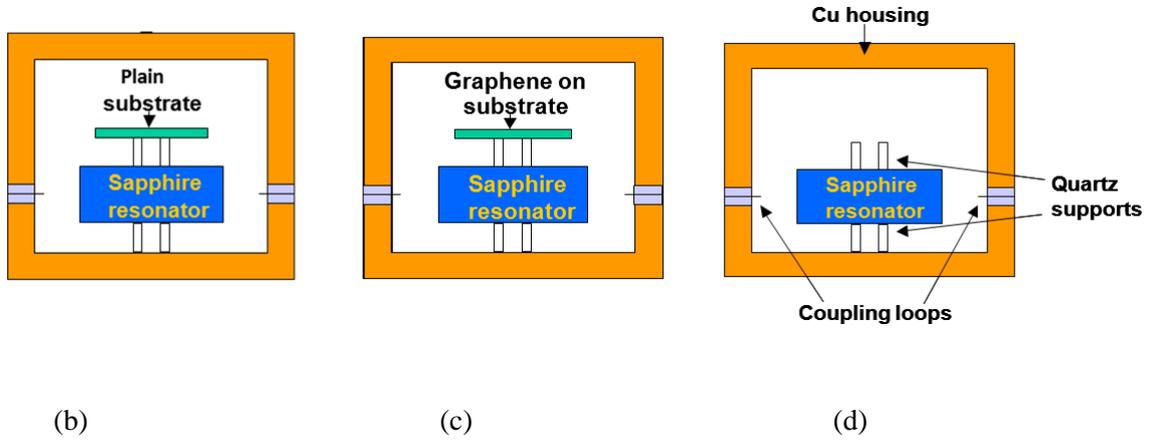

*Figure 1.* (a) Photo of sapphire puck and quartz spacer tubes inside copper housing (lid removed).( b-d) Schematic diagram of the high-Q sapphire dielectric resonator for measurement of the surface impedance of graphene samples.( b) a plain substrate ( c) an identical substrate with graphene film and (d) with neither substrate, just the dielectric resonator and support structures.

where $E$ is the field within the substrate and the integral is over the substrate volume. $W$ is the total stored energy in the puck and substrate system. $\varepsilon'_s$ is real part of the substrate permittivity. Due to the assumptions above that the graphene layer is very thin, its presence in the second measurement will not significantly perturb the total field distribution in the system. However the complex permittivity of the thin film may contribute significantly to the frequency shift and the linewidth shift

$$\Delta f_g = f_0 \left( \frac{(\varepsilon'_s - 1)\int E^2 dV + (\varepsilon'_g - 1)\int E^2 dv}{W} \right) \quad (2)$$

where now the second integral is over the volume of the graphene film only, and $\varepsilon'_g$ is real part of the permittivity of the graphene. Since the volume of the graphene is always far smaller than the substrate volume whereas the real permittivities are likely to be similar in magnitude the second term in the above integral can be ignored to first order.

Similar expressions can be written for the linewidth shifts in both cases

$$\Delta w_s = f_0 \left( \frac{\varepsilon''_s \int E^2 dV}{W} \right) \tag{3}$$

$$\Delta w_g = f_0 \left( \frac{\varepsilon''_s \int E^2 dV + \varepsilon''_g \int E^2 dv}{W} \right) \tag{4}$$

where $\varepsilon''_s$ and $\varepsilon''_g$ are imaginary part of the permittivity of the substrate and the graphene, respectively. Note that in the case of the imaginary components we cannot ignore the second term in the integral since graphene is a conductor with conductivity comparable to that of a metal, in which case $\varepsilon''_g \gg \varepsilon''_s$.

The readily measurable quantities are $\Delta f_s$ and $(\Delta w_g - \Delta w_s)$ and the latter may be expressed as

$$\Delta w_g - \Delta w_s = f_0 \left( \frac{\varepsilon''_g \int E^2 dv}{W} \right) = \varepsilon''_g \frac{\Delta f_s t_g}{(\varepsilon'_s - 1) t_s} \tag{5}$$

Here we have assumed that both for graphene and bare substrate the electric field is uniform throughout the thickness, a reasonable assumption provided the substrate is much thinner than the height of the puck. The aim of this measurement is to derive a surface resistance (or sheet resistance) $R_s$ value for the graphene film since this will also enable us to estimate the graphene conductivity (and the mobility if we are able to estimate the carrier density). We know that

$$R_s = \frac{1}{\sigma t_g} \tag{6}$$

There is a simple relationship between conductivity $\sigma$ and imaginary component of the dielectric constant $\varepsilon_g''$:

$$\sigma = 2\pi f_0 \varepsilon_0 \varepsilon''_g = \frac{2\pi f_0 \varepsilon_0 (\Delta w_g - \Delta w_s)(\varepsilon'_s - 1) t_s}{\Delta f_s t_g} \tag{7}$$

So finally our expression for $R_s$ becomes independent of graphene thickness $t_g$

$$R_s = \frac{\Delta f_s}{2\pi\, f_0 \varepsilon_0 \left(\Delta w_g - \Delta w_s\right)\left(\varepsilon_s' - 1\right) t_s} \qquad (8)$$

## 4. Comparisons of measurements on CVD and GO graphene

The $TE_{011}$ resonance in the sapphire dielectric puck occurs at around 10.3 GHz. The presence of a 10x10 mm bare quartz substrate, placed directly on top of the sapphire resonator, shifts the frequency (downwards) by around 200MHz while producing no significant reduction in the quality factor ($Q$) of the resonance, which at room temperature is around $1\times10^4$. The presence of a single layer of graphene on such a quartz substrate produces a large reduction in $Q$ value and a further small shift in the resonant frequency. In figure 2 the upper trace shows an example of a resonance with plain quartz substrate and the lower trace shows a similar trace with a sample of CVD grown graphene [12] transferred onto a nominally identical quartz substrate. Note the reduction $Q$ by a factor of approximately 10 in the latter case.

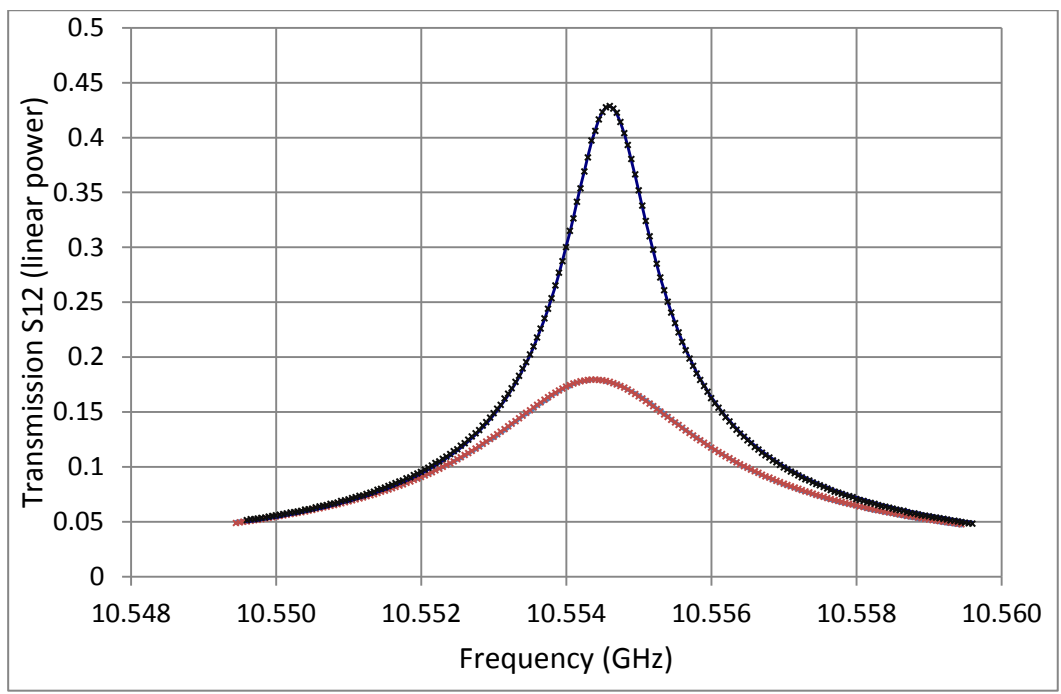

*Figure 2. Measured $S_{12}$ transmission versus frequency for both a bare quartz substrate (upper trace) and a quartz substrate with transferred CVD graphene (lower trace). The narrower resonance (higher Q) corresponds to the bare quartz sample and note that the linewidth is approximately 3 times greater for the CVD sample compared with bare quartz. Crosses are experimental points and solid lines are the Lorentzian fit to the data ( NB the two are almost indistinguishable to the eye).*

The resonant frequency and linewidth are measured in transmission with a vector network analyser (VNA) being used to measure $S_{12}$ as a function of frequency. The internal software of the VNA (HP 8720) is used to collect the centre frequency and 3dB linewidth but in addition the full trace data

(200 frequency points) is downloaded to a computer and a non-linear least squares fit routine is used to fit the data to a skewed Lorentzian lineshape. By this method we have found that the uncertainty in centre frequency and in linewidth is reduced by a factor of 10, compared with the results output directly from the VNA.

Using the above dielectric resonator technique we have made measurements on a number of samples of graphene grown by different techniques. For samples with rather high conductivity it has sometimes been necessary to reduce the influence of the graphene layer on the quality factor of the system by inserting a small quartz tube spacer a few mm in height, between the sapphire puck and the graphene sample (as shown schematically in figure 1). In this way the influence on the resonant properties is reduced so that the $Q$ value can be accurately measured.

In Table 1 and table 2 we compare liquid-phase grown graphene oxide (GO), subsequently reduced (rGO) [15], CVD graphene grown on a copper catalyst layer [12], and then transferred to a clean quartz substrate (permittivity 4.4 , 10x10mm and 0.5mm thick). Finally two separate samples of epitaxially grown graphene on SiC are also shown. These are of considerably higher quality than any of the other films examined, as shown by the much lower sheet resistance values. Note that the range of sheet resistance values measured spans almost four orders of magnitude, demonstrating the great sensitivity and dynamic range of the method. In our method we can derive sheet resistance $R_s$ without the need to measure the thickness of the graphene since sheet resistance derived by equation (8) is independent of thickness.

It is clear from the results in the Table that, as would be expected, the rGO sample has a conductivity considerably higher than the as-grown GO sample for same thickness. It is the reduction process which converts the sample into a semi-metallic state. Comparing the rGO and CVD sample conductivities it is clear that the latter is more metallic than the former, again unsurprising given the nature of the wafer-scale growth process for CVD. For monolayer graphene on SiC the conductivity is more than two orders of magnitude greater than the monolayer CVD graphene and the sheet resistance is nearly three orders of magnitude lower than the CVD sample.

*Table 1.* *Summary of properties of monolayer CVD graphene, reduced graphene oxide on quartz substrates and monolayer graphene on SiC,*

| Sample | Graphene thickness (nm) | $f_0$ (GHz) | $\Delta f$ (MHz) | $\Delta w_g$-$\Delta w_s$ (MHz) | Conductivity $\sigma$ (S/m) | Sheet resistance $R_s(\Omega/\square)$ |
|---|---|---|---|---|---|---|
| 1-layer reduced GO | 0.4 | 10.5504 | 1.169 | 0.0243 | $4.82 \times 10^4$ | 48222 |
| 1-layer CVD | 0.4 | 10.4596 | 140.7 | 10.91 | $1.92 \times 10^5$ | 13038 |
| 1-layer on SiC (sample 1) | 0.4 | 10.5619 | 7.463 | 47.30 | $2.03 \times 10^7$ | 61.7 |
| 1-layer on SiC (sample 2) | 0.4 | 10.5600 | 5.057 | 73.38 | $4.64 \times 10^7$ | 26.97 |

*Table 2.* *Summary of properties of 5-layer Graphene oxide and reduced graphene oxide on quartz substrate.*

| Sample | Graphene thickness (nm) | $f_0$ (GHz) | $\Delta f$ (MHz) | $\Delta w_g - \Delta w_s$ (MHz) | Conductivity $\sigma$ (S/m) | Sheet resistance $R_s(\Omega/\square)$ |
|---|---|---|---|---|---|---|
| 5-layer reduced GO | 2.0 | 10.5502 | 122.8 | 0.083 | $3.38 \times 10^4$ | 14788 |
| 5-layer Graphene oxide | 2.0 | 10.5502 | 1.165 | 0.019 | $8.30 \times 10^3$ | 60277 |

## 5. Conclusions

The analysis presented above, and the results shown in Table 1 and 2, confirm that the dielectric resonator technique provides a quick and straightforward method for analysing the conducting properties of graphene samples. Note that no patterning or electrical contacts are required, which may damage or compromise the sample quality. The sample can be placed on top of the puck or quartz spacer without requiring adhesive, another potentially damaging addition. The method has the great sensitivity and dynamic range. For single-layer conductivity of CVD sample is $1.92 \times 10^5$ S/m which is about 5 times better that the reduced GO. The best conductivity ($4.64 \times 10^7$ S/m) and sheet resistance $R_s$ (27 $\Omega/\square$) are for single-layer graphene on SiC. Thus the method shows great promise for rapid quality control and characterisation. It also may be extended in future to cryogenic or elevated temperature measurement [16].

**Acknowledgments:**

We thank Dr. C. Mattevi of Imperial College for providing graphene oxide samples. This work was funded by the UK NMS Programme and the EU EMRP Project MetNEMS (NEW-08). The EMRP is jointly funded by the EMRP participating countries within EURAMET and the European Union.